\documentclass{moriond}

\def\be{\begin{equation}}
\def\ee{\end{equation}}
\def\bea{\begin{eqnarray}}
\def\eea{\end{eqnarray}}

\newcommand{\eqn}{equation}
\newcommand{\lb}{\left (}
\newcommand{\rb}{\right )}
\newcommand{\GeV}{{\ensuremath\rm GeV}}
\newcommand{\TeV}{{\ensuremath\rm TeV}}

\newcommand{\Photo}{\includegraphics[height=35mm]{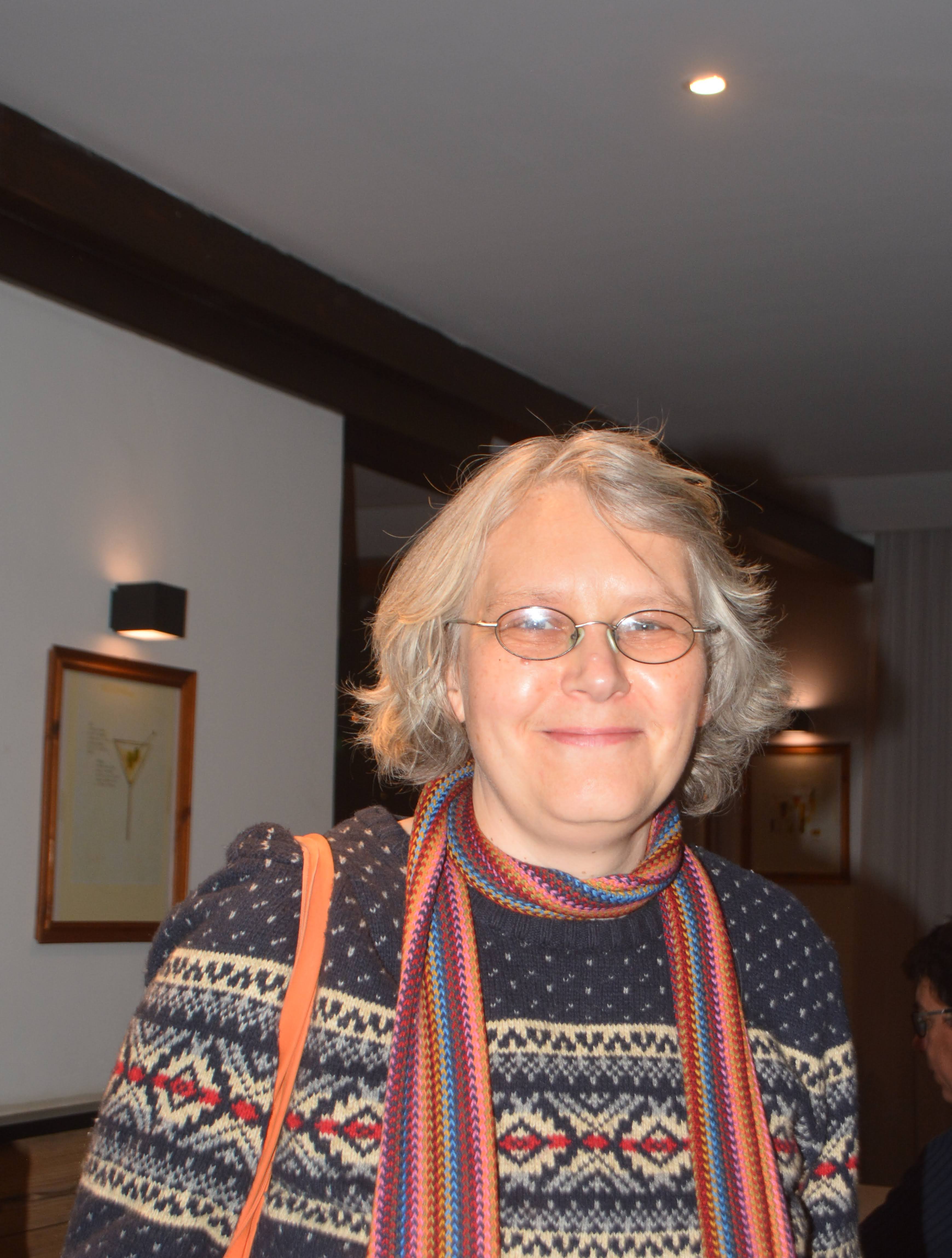}}

\begin{document}
\vspace*{4cm}
\title{Extended scalar sectors from all angles (in 15 minutes)}

\author{ T. Robens }

\address{Institute Rudjer Boskovic,\\
  Bijenicka cesta 54, 10000 Zagreb, Croatia\\ and\\
Theoretical Physics Department, CERN,\\
 1211 Geneva 23, Switzerland}

\maketitle\abstracts{
In this proceedings contribution, I briefly summarize various aspects that are important in the discussions of new physics searches with novel scalar states, at current and future colliders. In particular, I give a brief glance on the status of two Higgs doublet models, and discuss multi-scalar production as well as interference effects in Di-Higgs searches. I also mention searches of new scalar final states at possible Higgs factories.\\
RBI-ThPhys-2025-21, CERN-TH-2025-083}


\section{Introduction}
Since the discovery of a particle that complies with the properties of the Higgs boson predicted by the Standard Model (SM) (see e.g. \cite{ATLAS:2022vkf,CMS:2022dwd} for the Run 1 legacy papers), particle physics has entered an exciting era. One important question is whether the boson discovered by the LHC experiments is indeed the Higgs boson of the Standard Model, or whether it is part of an enhanced scalar sector the SM is embedded in.

Currently, both theoretical and experimental uncertainties still allow for additional particle content without violating current observations. However, in the construction of new physics scenarios  a number of constraints need to be obeyed. From the theoretical side, these include checks that the potential is bounded from below, stays unitary (typically tested via perturbative unitarity (PU)), and that couplings themselves remain perturbative up to a certain scale. Furthermore, the model predictions should allow to describe current data, in particular the current findings for the 125 \GeV~ Higgs-like scalar, and should not be in conflict with null-results from searches. Important constraints can also stem from electroweak precision observables (EWPOs). For models that contain a dark matter candidate in addition bounds from astrophysical findings, as e.g. the measurement of relic density or direct detection constraints, have to be obeyed.

In general, such novel physics scenarios can manifest themselves in various ways: {\sl (a)} they can lead to direct observations of new resonances, ideally in an invariant mass distribution; {\sl (b)} they can lead to novel signatures that do not exist in the Standard Model, e.g. events with large missing energy; {\sl (c)} they can impact observables at higher orders, e.g. through one-loop contributions. In the following, I will mainly concentrate on installments of the first category.
\section{Two Higgs Doublet Models (2HDMs)}
A popular new physics extension are models with a second $SU(2)\,\times\,U(1)$ scalar doublet. Such models (see e.g. \cite{Branco:2011iw} for a concise overview) come with several new scalar states. The scalar content is given by
\begin{\eqn}
h,\,H,\,A,\,H^\pm,
\end{\eqn}
where one of the two CP-even neutral scalars $h,\,H$ has to comply with the measurements of the 125 \GeV~ resonance observed by the LHC experiments \cite{ATLAS:2022vkf,CMS:2022dwd}. In general, such models can give rise to flavour changing neutral currents, which can be supressed by imposing appropriate additional symmetries on the Lagrangian. Furthermore, it is convenient to chose the masses of the scalars as input variables, together with the sum of the squared vacuum expectation values (vevs) $v^2\,=\,v_1^2+v_2^2$, the ratio of these vevs $\tan\beta\,\equiv\,\,\frac{v_2}{v_1}$, a combination of mixing angles $\cos\lb \beta-\alpha\rb$, as well as an additional potential parameter $m_{12}$.

\begin{center}
\begin{figure}
\begin{center}
\includegraphics[width=0.49\textwidth]{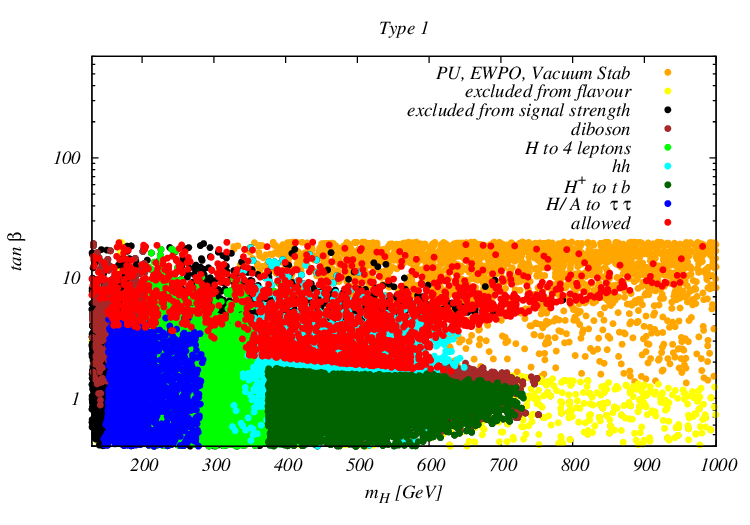}
\includegraphics[width=0.49\textwidth]{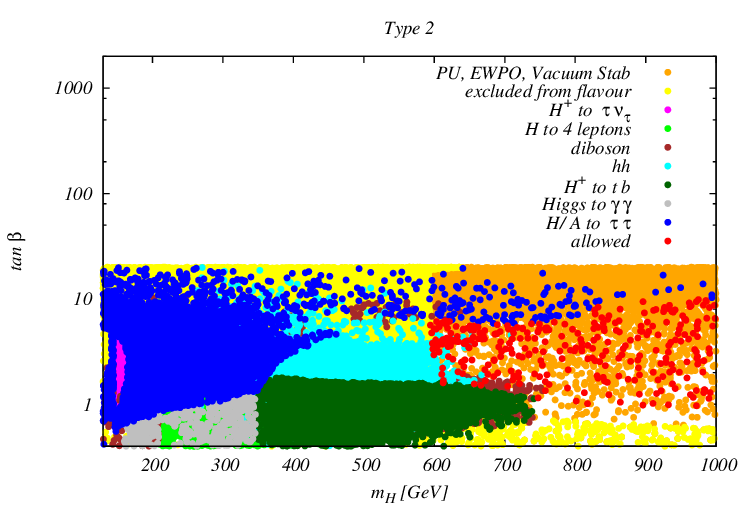}
\caption{\label{fig:2hdms} Current constraints on 2HDMs using bounds available within thdmtools and HiggsTools. All heavy masses are set equal, all other parameters are floating. See text for details.}
\end{center}
\end{figure}
\end{center}

As all new physics scenarios, 2HDMs are subject to a large number of theoretical and experimental constraints. Many tools are currently on the market to test the available parameter space against such constraints, as e.g. 2HDMC \cite{Eriksson:2009ws,Eriksson:2010zzb}, ScannerS \cite{Coimbra:2013qq,Muhlleitner:2020wwk}, HiggsTools \cite{Bahl:2022igd}, and SuperIso \cite{Arbey:2018msw}. In figure \ref{fig:2hdms}, I show the results of a recent scan of the 2HDM parameter space in a specific configuration where all additional masses are set equal, but all other parameters are allowed to float. I display two cases where either all fermions couple to the second doublet $\Phi_2$, labelled type 1, or where instead the down-type fermions couple to $\Phi_1$, labelled type 2. In particular it becomes clear that the different coupling structures lead to large differences in the allowed parameter space. A main reason for this are the flavour contraints from $B\,\rightarrow\,X_s\,\gamma$ that require a mass $m_{H^\pm}\,\geq\,600\,\GeV$ \cite{Misiak:2017bgg,Misiak:2020vlo} in type 2 scenarios. The plots have been obtained using thdmtools \cite{Biekotter:2023eil}. Experimental constraints from collider searches and measurements are included using HiggsTools \cite{Bahl:2022igd}. The list of contributing experimental channels can be found in the appendix of \cite{mytalk}.

\section{Multi-scalar final states}
Of similar interest are processes with multi-scalar final states. I here focus on processes with 2 additional neutral CP-even scalar bosons, as e.g. realized in a simple two real singlet extension model (TRSM) \cite{Robens:2019kga,Robens:2022nnw}. In this scenario, the following processes are of particular interest: 
\begin{\eqn}
p\,p\,\rightarrow\,h_3\,\rightarrow\,h_1\,h_2;\;p\,p\,\rightarrow\,h_i\,\rightarrow\,h_j\,h_j,
\end{\eqn}
where we assume a mass ordering such that $M_1\,\leq\,M_2\,\leq\,M_3$ and where in the second process none of the particles corresponds to the 125 \GeV~ resonance. These processes within the TRSM  and the corresponding benchmark planes (BPs) were first discussed in \cite{Robens:2019kga} and then updated in \cite{Robens:2022nnw,Robens:2022hvu,Robens:2023pax,Robens:2023oyz}.

\begin{center}
\begin{figure}
\begin{center}
\includegraphics[width=0.4\textwidth]{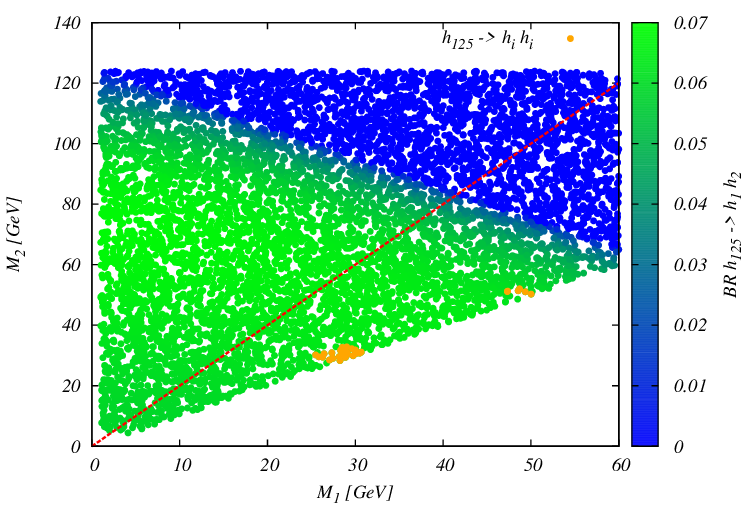}
\includegraphics[width=0.4\textwidth]{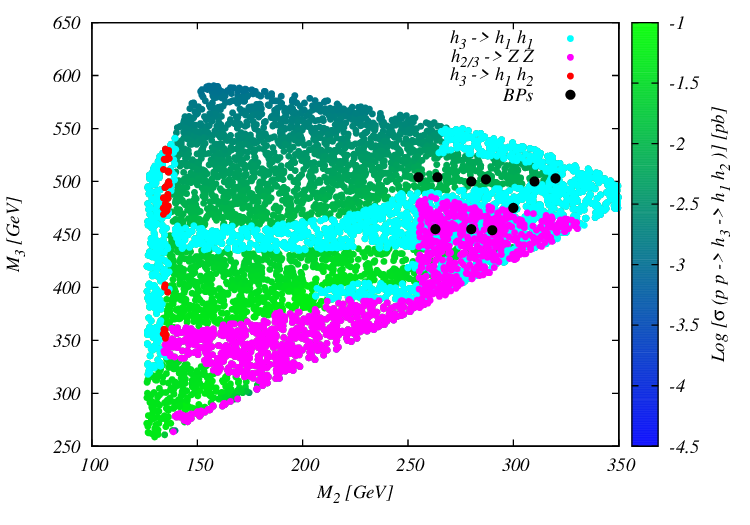}\\
\includegraphics[width=0.4\textwidth]{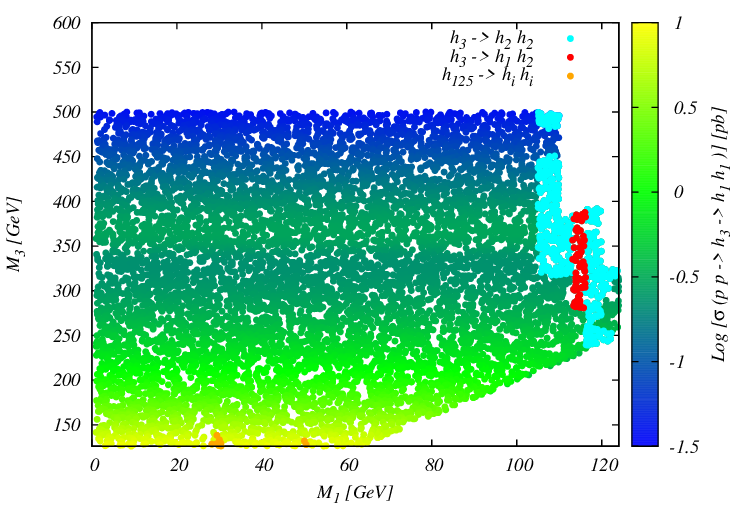}
\includegraphics[width=0.4\textwidth]{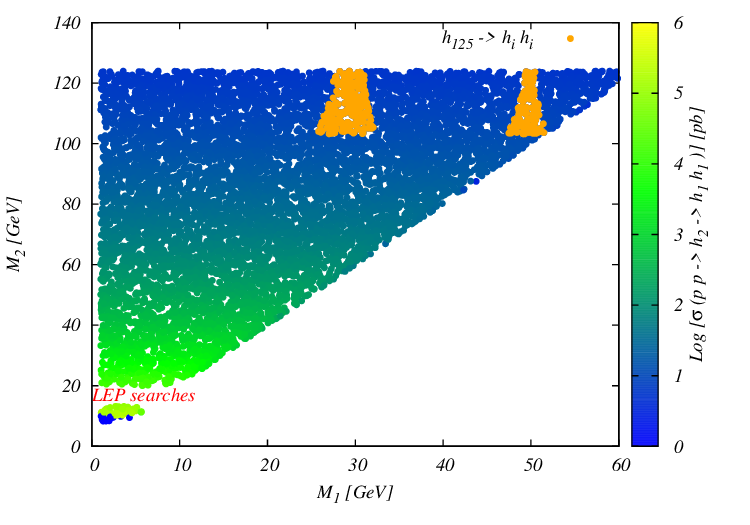}
\end{center}
\caption{\label{fig:trsm} Updated benchmark planes for the TRSM in the respective di-mass planes: asymmetric decays with BP1 and BP3 {\sl (top row)} as well as symmetric decays for BP5 and BP4 {\sl (bottom row)}. Shown is the maximal decay for the 125 \GeV~ resonance for BP1 as well as production rates for the other three benchmarks at a 13 \TeV~ LHC. See text for details.}
\end{figure}
\end{center}

Figure \ref{fig:trsm} shows the current benchmark planes for antisymmetric/ symmetric decays taking the most recent constraints as given by HiggsTools into account. In particular, now searches for $h_{125}\,\rightarrow\,h_i\,h_i$ \cite{CMS:2024uru}, $h_3\,\rightarrow\,h_{i}\,h_{i}$ \cite{ATLAS:2022xzm}, $h_3\,\rightarrow\,Z\,Z$ \cite{ATLAS:2020tlo} and $h_3\,\rightarrow\,h_1\,h_2$ \cite{CMS:2021yci} with respect to previous publications are included.
\section{Interference effects in Di-Higgs searches}
Another important topic to discuss are interference effects for processes that can be mediated by an additional heavy scalar resonance. I here concentrate on processes leading to di-Higgs final states. The corresponding resonance contribution is known to interfere with both the SM-like off-shell s-channel resonance as well as continuum contributions from the top-box mediated process, see e.g. \cite{DiMicco:2019ngk} for an overview and further references.

\begin{center}
\begin{figure}
\begin{center}
 \includegraphics[width=0.35\textwidth]{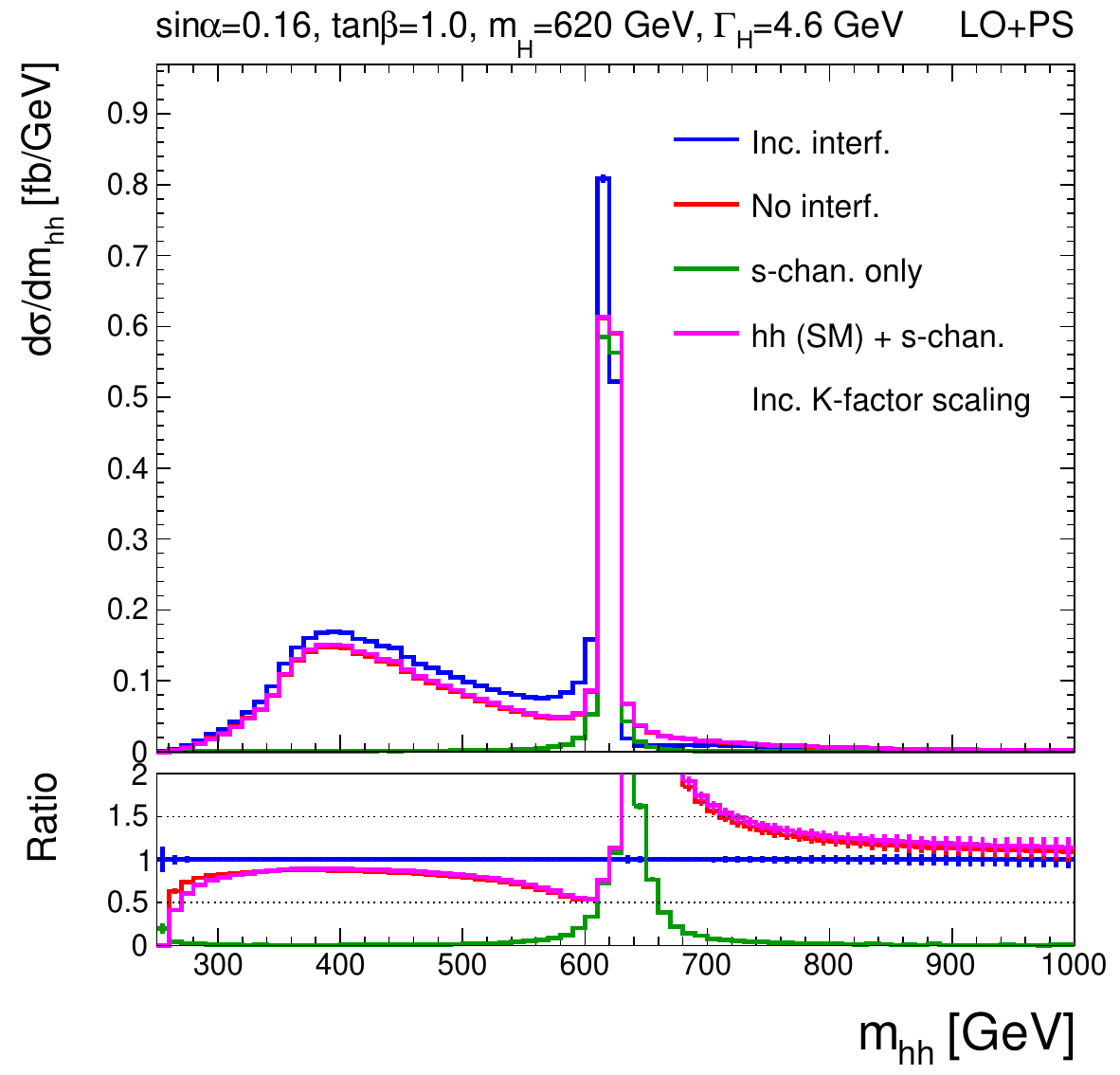}
  \includegraphics[width=0.35\textwidth]{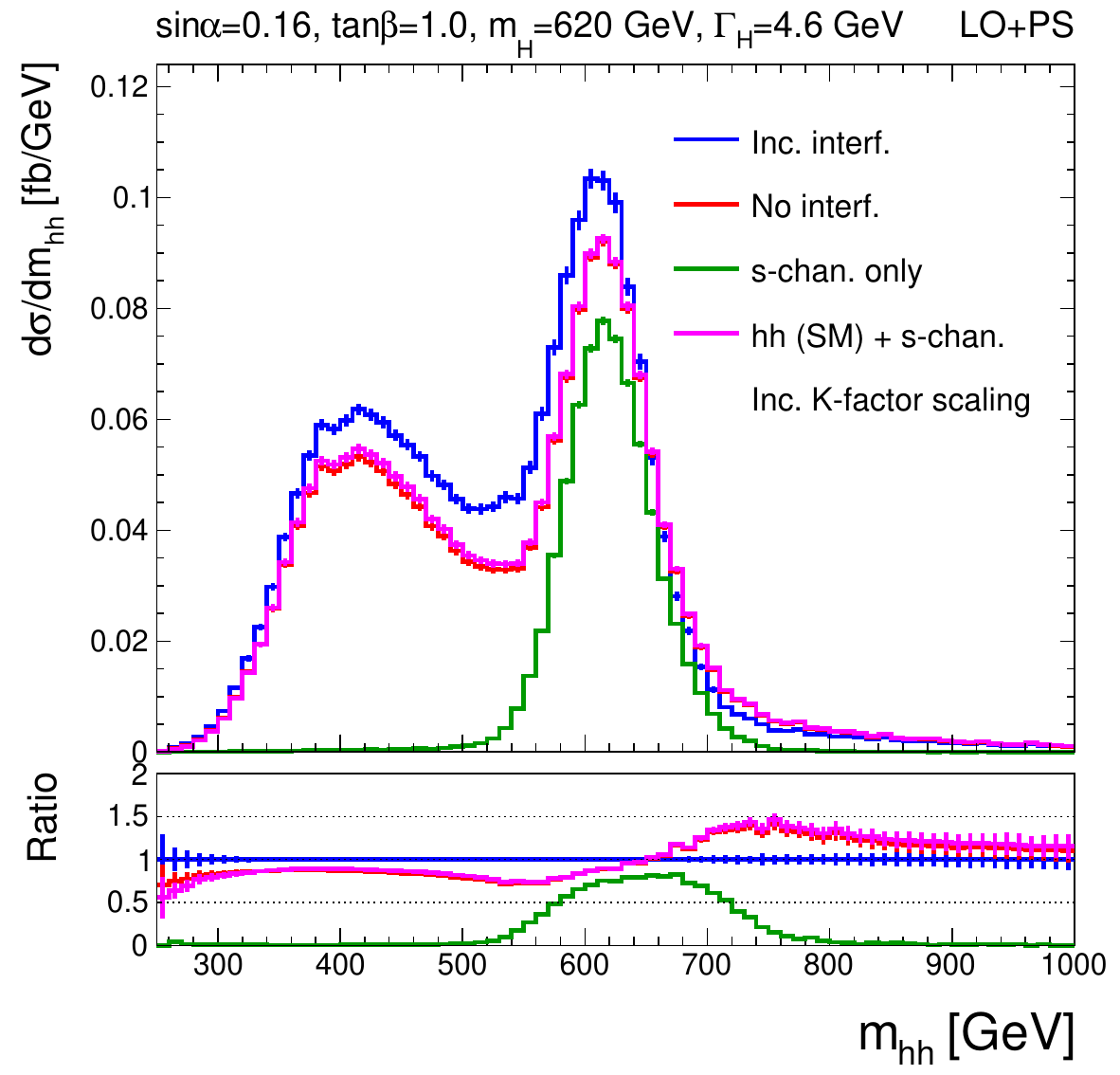}\\
\includegraphics[width=0.35\textwidth]{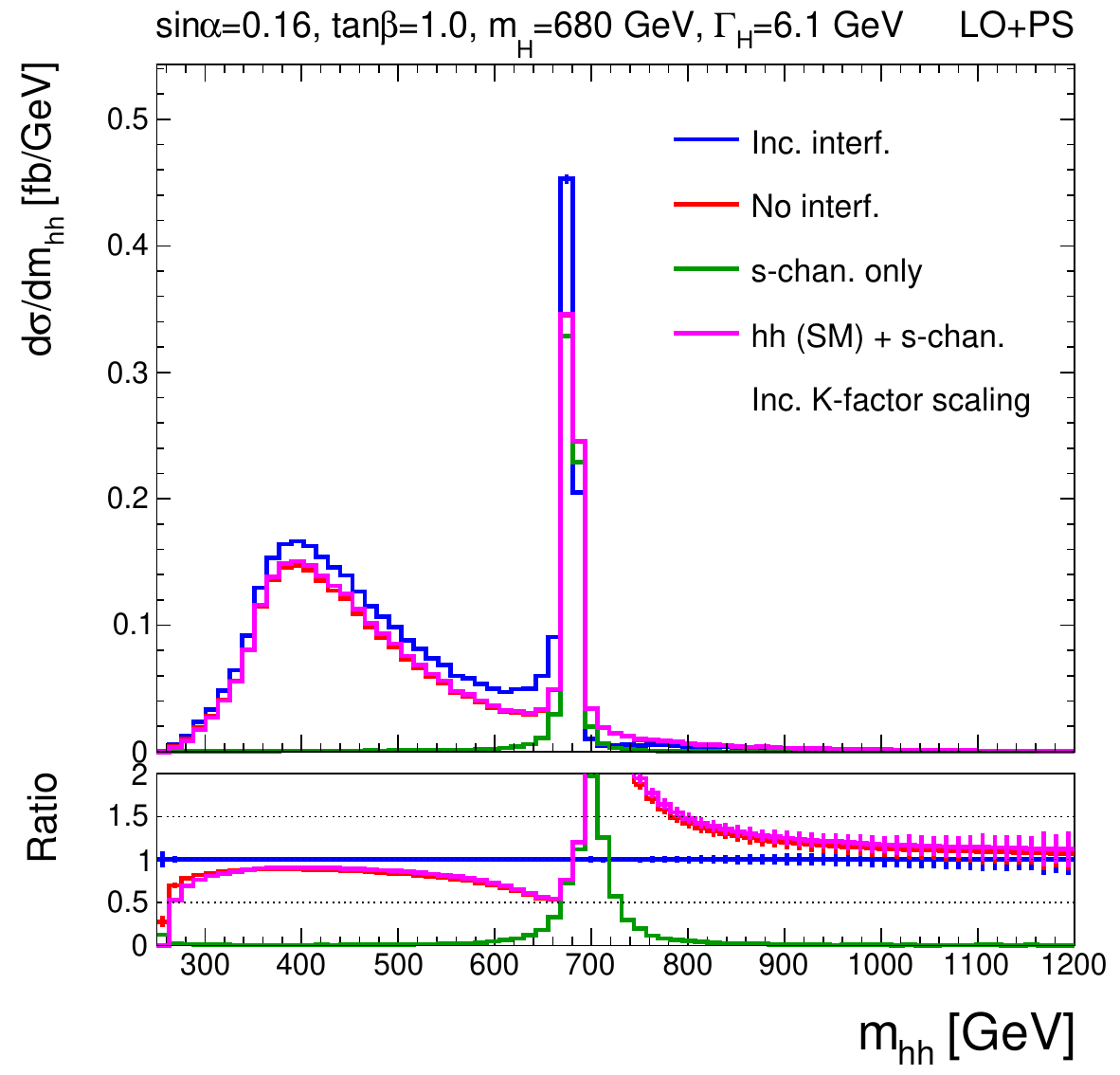}
  \includegraphics[width=0.35\textwidth]{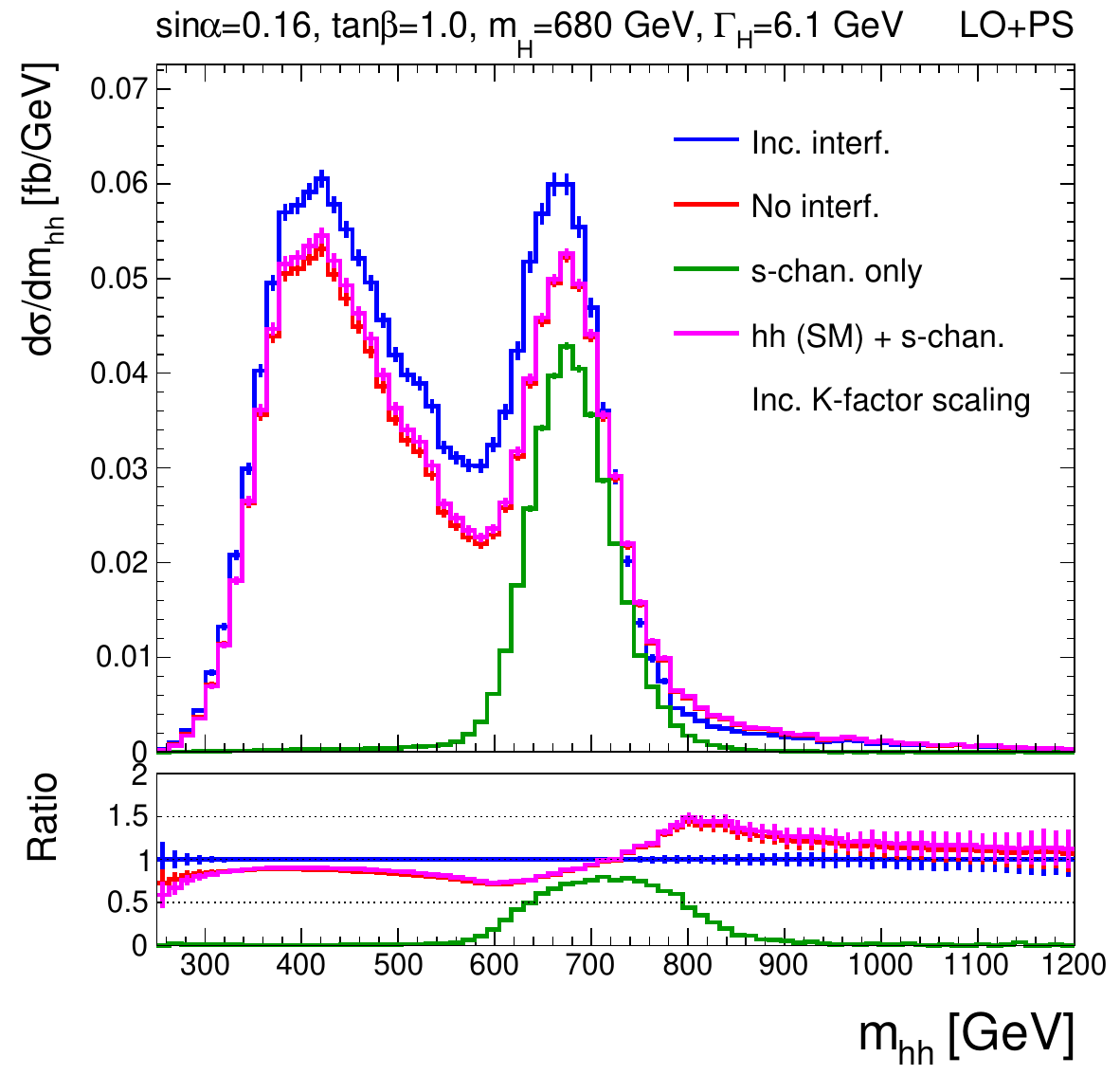}
\end{center}
\caption{\label{fig:interf} Di-Higgs invariant mass distributions from different contributing terms, for two different scenarios, before {\sl (left)} and after {\sl (right)} smearing. See text for details as well as description of different contributions.}
\end{figure}
\end{center}

A most realistic scenario would be the simulation of each possible benchmark point, defined by a resonance mass, a supression factor/ mixing angle, and the width of the heavy resonance, for various choices of the latter two quantities. However, a detailed simulation of such signals is time consuming. In \cite{Feuerstake:2024uxs}, we therefore propose a reweighting tool that can be used to reweight one time generated samples using different weights that reflect different choices for these input parameters. We also showcase several scenarios where such effects can be large. Two of these are shown in figure \ref{fig:interf}, taken from \cite{Feuerstake:2024uxs}.

In this figure, we display the full process {\sl (blue)}, the contribution stemming from the heavy resonance only {\sl (green)}, the contributions without interference {\sl (red)}, as well as the SM-like contribution and the heavy resonance contribution added without interference terms {\sl (magenta)}. Shown are the di-Higgs invariant mass distributions before and after smearing,({\sl (left)} and {\sl (right)}, respectively), for two different scenarios in the top and bottom row. It is clear that the SM background cannot be neglected in such searches, and that there can be large contributions from the interference terms.
\section{Scalars at Higgs factories}
\begin{center}
\begin{figure}
\begin{center}
\includegraphics[width=0.65\textwidth]{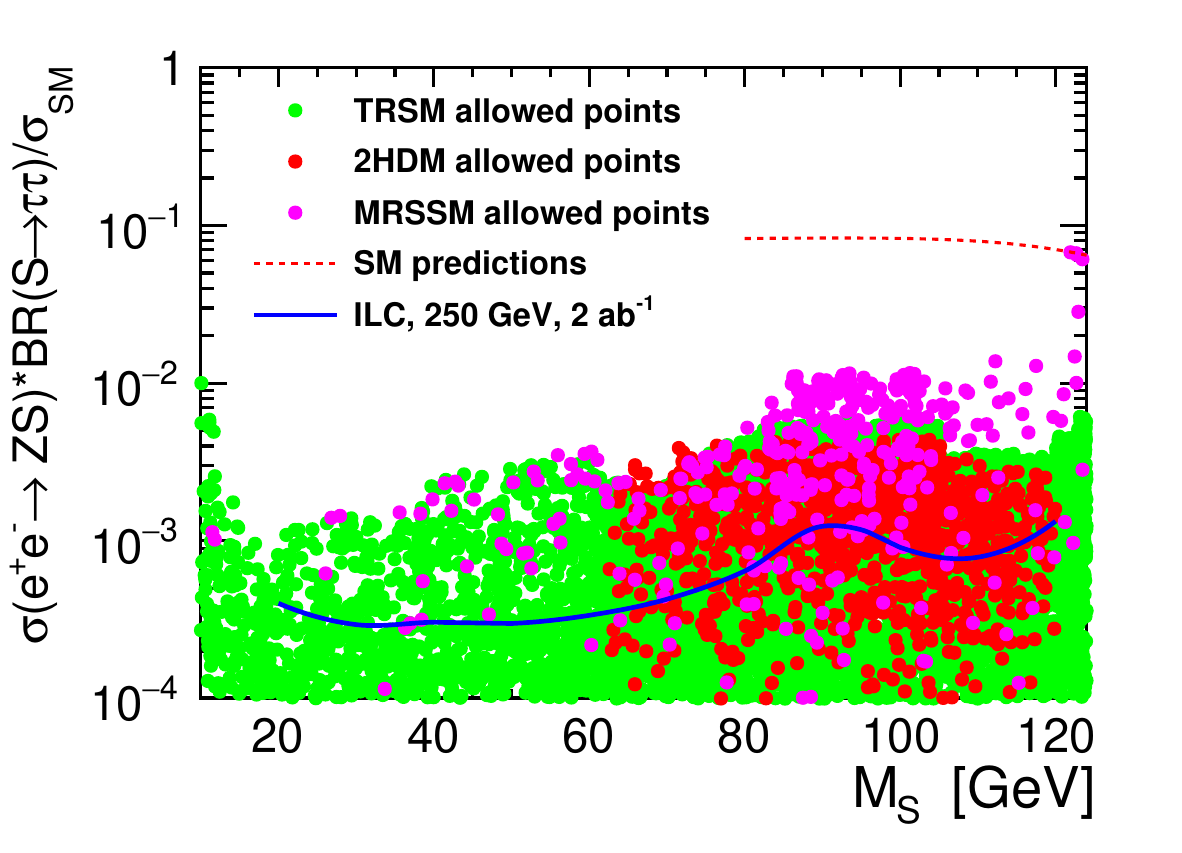}
\caption{\label{fig:higgsfac} Feasibility study for scalar-strahlung with di-tau final states as a function of the mass, as well as possible model parameter points that can render this final state.}
\end{center}
\end{figure}
\end{center}

Finally, we turn to searches for extra scalars at Higgs factories, i.e. $e^+\,e^-$ colliders with center-of-mass energies of around $240\,-\,250\,\GeV$ as a prime target. These processes have now been discussed in a large number of documents, see e.g. \cite{Robens:2022zgk,deBlas:2024bmz,Robens:2024wbw,LinearCollider:2025lya} for recent work. In particular, in \cite{deBlas:2024bmz} several processes were identified as target processes that should be investigated in order to identify important benchmark scenarios and feasibilities that could serve as input for the current European Strategy for Particle Physics. An important process to consider is the so-called scalar-strahlung, where a new physics scalar is emitted from a Z-boson in the s-channel. The new physics scalar can then decay into various final states. In figure \ref{fig:higgsfac}, we show an example for this feasibility study for di-tau final states, where also possible new physics scenarios are shown that can provide rates that are accessible at such a collider. The figure is taken from \cite{Robson:2920434}.
\section{Conclusions}
In this work, I briefly touched upon various aspects of searches for novel scalar states at current and future colliders. I mentioned an example of current allowed parameter space for a popular new physics extension, a two Higgs doublet model, as well as multi-scalar final states. I also emphasized the importance of including interference effects in searches for di-Higgs final states. Finally, I mentioned possibilities to investigate new scalars at future Higgs factories.

Apart from a pragmatic approach, new scalar states can also be useful in order to solve important open issues of the SM. One popular example, although not mentioned here, is the stabilization of the vacuum by additional scalar content or the possibility of a strong first-order electroweak phase transition. Any search for or discovery in a new physics channel  is therefore well motivated and will allow to further understand the history as well as constituents of the universe.

\section*{Acknowledgments}
The author wants to thank the organizers for the invitation, financial support, and the revival of my skiing skills. This work is furthermore supported by the Croatian Science Foundation (HRZZ) under project IP-2022-10-2520.
\section*{References}
\bibliography{moriond}

\end{document}